\documentclass[prl,twocolumn,aps,longbibliography]{revtex4}
\usepackage{bm}
\usepackage{graphicx}
\usepackage{amssymb}
\usepackage{amsmath}
\usepackage{eufrak}
\usepackage{color}
\usepackage[utf8]{inputenc}
\usepackage[T2A]{fontenc}
\usepackage{ulem}
\usepackage[unicode=true,colorlinks=true,citecolor=blue,urlcolor=blue]{hyperref}

\renewcommand{\Re}{\mathop{\rm Re}}
\renewcommand{\Im}{\mathop{\rm Im}}
\newcommand{\Tr}{\mathop{\rm Tr}}

\newcommand{\e}{\mathrm{e}}
\renewcommand{\i}{{\rm i}}

\renewcommand{\emph}{\textit}

\newcommand{\Js}{{\bm{\mathcal{J}}}}
\newcommand{\Jdr}{{\bm{\mathcal{J}}}_{\rm dr}}
\newcommand{\Jdiff}{{\bm{\mathcal{J}}}_{\rm diff}}

\newcommand{\nix}[1]{}

\begin{document}

\title{
Electrical spin orientation, spin-galvanic and spin-Hall effects \\ in disordered two-dimensional systems
}
\author{D.\,S. Smirnov}
\email[Electronic address: ]{smirnov@mail.ioffe.ru}

\author{L.\,E.~Golub}
\affiliation{Ioffe Institute, 194021 St.~Petersburg, Russia}

\begin{abstract}
In disordered systems, the hopping conductivity regime is usually realized at low temperatures where 
spin-related phenomena differ strongly from the case of delocalized carriers. We develop the unified microscopic theory of current induced spin orientation, spin-galvanic and spin-Hall effects for the two-dimensional hopping regime. We show that the corresponding susceptibilities are proportional to each other and determined by the interplay between the drift and the diffusion spin currents. Estimations are made  for realistic semiconductor heterostructures using the percolation theory. We show that the electrical spin polarization in the hopping regime increases exponentially with increase of the concentration of localization sites and may reach a few percents at the crossover from the hopping to the diffusion conductivity regime.
\end{abstract}


\maketitle{}

\textit{Introduction.}---Spin physics is a rapidly growing area of research in condensed matter science aimed at
 the creation, manipulation and detection of spins in various systems~\cite{Speculation1}. Important and fundamentally interesting  results, also promising for possible future applications, have been obtained in semiconductors and semiconductor nanostructures~\cite{Dyakonov_book}.
The cornerstones in semiconductor spintronics are spin orientation, spin transfer and spin readout. The remarkable progress has been achieved in 
last-decade experiments
in all three directions including ultrafast optical spin injection~\cite{Ultrafast1,Orientation1}, low-dissipation spin current manipulation~\cite{SpinCurrent}, and nearly non-destructive spin measurements~\cite{Measurement1}. 
%
A challenging problem in the spin physics is how to affect the spin by instantaneous non-magnetic methods, in particular, by electric fields~\cite{Electrical1,Electrical2}. 
%
The key to the electrical spin control is the spin-orbit interaction~\cite{Manchon2015}, which linearly couples spin and momentum components of carriers. It allows for the current-induced spin polarization (CISP) --- a phenomenon where the electric current flow is accompanied by a homogeneous orientation of carrier spins. 
This problem is mostly studied in semiconductors, see Ref.~\cite{Ganichev_Handbook} for review. Recent progress in the field is related to precise electrical control of spin in semiconductor epilayers~\cite{Sih_2014,Beschoten}. The problem of CISP in two-dimensional~(2D) semiconductor heterostructures is investigated theoretically in detail, 
including nonlinear regimes of CISP~\cite{NOVA,Vignale_PRB_2016}.

There are two more phenomena closely related to CISP. The first one is a generation of electric current in systems with a nonequilibrium spin polarization referred to as the spin-galvanic effect (SGE)~\cite{Ganichev_Nature_2002}. SGE has been studied in various 2D semiconductor systems where nonequilibrium spin polarization has been created by means of optical excitation~\cite{Spin_review}. One more phenomenon is the spin-Hall effect (SHE) consisting in a generation of the spin current in the presence of the electric current~\cite{Dyakonov_book}. All three effects are phenomenologically introduced as follows:
\begin{equation}
	\bm s = \hat{\bm \sigma}_\text{CISP} \bm E,
	\quad
		\bm j = \hat{\bm \sigma}_\text{SGE} \bm s,
	\quad
		\bm{\mathcal J} = \hat{\bm \sigma}_\text{SHE} \bm E.
\label{eq:3effects}
\end{equation}
Here $\bm s$ is the nonequilibrium spin polarization, $\bm E$ and $\bm j$ are the electric field and electric current density, and $\bm{\mathcal J}$ is the component of the spin current describing
the flux density of spins oriented along the normal to the 2D plane.

Despite a deep investigation of CISP, SGE and SHE, all previous activities were devoted to 
delocalized electrons, which weakly feel the static disorder as a source of rare momentum scattering. However, the role of the disorder is drastically enhanced
at low temperatures when carriers are localized in minima of potential energy.
In contrast to free electron systems, the localized carriers preserve their spin coherence for hundreds of nanoseconds due to suppression of the Dyakonov-Perel spin relaxation mechanism~\cite{KKavokin-review}. 
The record spin coherence times have been demonstrated for semiconductor quantum dot structures~\cite{LongT1,LongT2,LongT3}. For this reason, the spin properties of localized electrons attract rapidly growing attention. Application of electric field to such systems induces directed hops of electrons between localization sites, so-called \textit{hopping conductivity regime}. Spin relaxation~\cite{Intronati}, spin dynamics~\cite{Burkov2003,Shumilin}, spin noise~\cite{Glazov2015,Glazov2016} and \textit{ac} spin Hall effect~\cite{KozubPRL} have been recently studied in the hopping regime.
However neither CISP, nor SGE, nor \textit{dc} SHE have been considered. In this Letter, we fill this gap and 
describe the effects of the spin, electric current and spin current mutual conversion in the hopping regime.

\begin{figure}[t]
\centering\includegraphics[width=\linewidth]{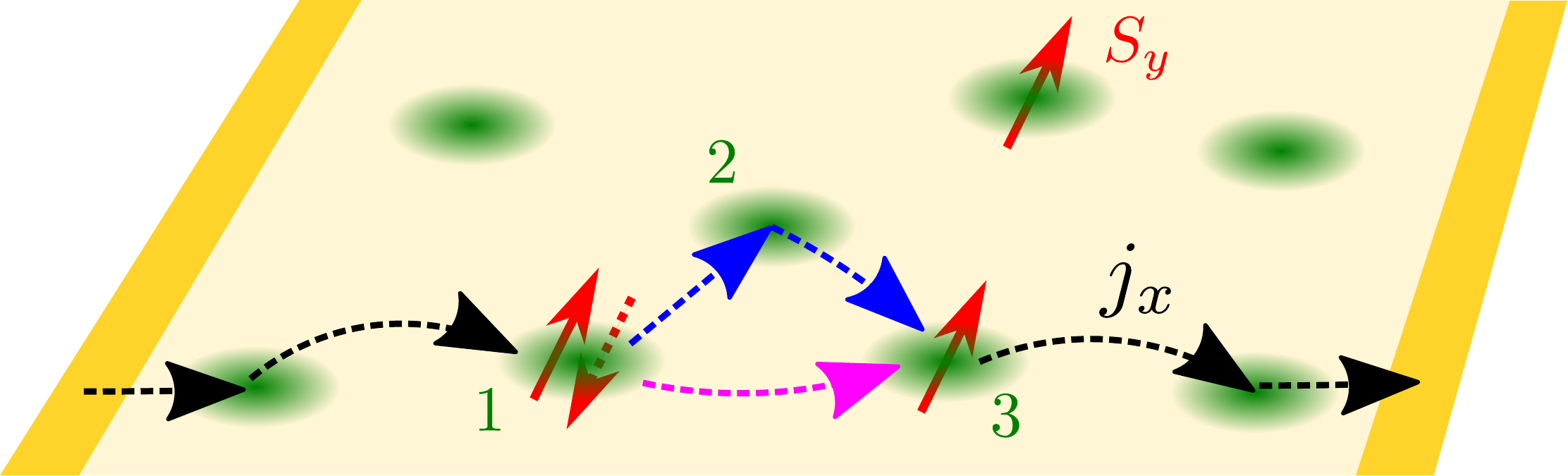}
\caption{ Illustration of CISP in the 2D hopping regime: electrons rarely hop between localization sites (green areas). In the presence of electric current, $j_x$, the quantum interference between the direct and indirect hopping paths shown respectively by magenta and blue arrows leads to spin polarization, $S_y$.}
\label{fig:sketch}
\end{figure}

\textit{Model.}---The effective electron Hamiltonian describing spin-orbit interaction in 2D heterostructures grown along $[001]$ direction has the form
\begin{equation}
\label{H_SO}
	{\cal H}_\text{SO} = \beta_{\mu\nu}\sigma_\mu k_\nu = \beta_{xy} \sigma_x k_y + \beta_{yx} \sigma_y k_x .
\end{equation}
Here $x \parallel [1\bar{1}0]$ and $y \parallel [110]$ are the coordinates in 2D plane, $\sigma_{x,y}$ are the Pauli matrices,
$\bm k = -{\rm i}\bm\nabla$, and $\beta_{\mu\nu}$ are spin-orbit constants caused by both bulk- and structure-inversion asymmetry~\cite{Spin_review,QW110}. We consider a 2D ensemble of electrons localized at random sites in a weak \textit{dc} electric field, see Fig.~\ref{fig:sketch}. The 2D concentration of carriers, $n$, is assumed to be much smaller, than the concentration of sites, $n_s$. This situation is realized, for example, in ensembles of weakly charged quantum dots or in $n$-doped QWs compensated by $p$ doping of barriers ($D^0$ or $D^-$ centers). Note that our theory can be equally applied to the ensembles of holes, but the electron tunneling between the sites is facilitated as compared with holes because the effective mass in the conduction band is as a rule smaller than that in the valence band.

A microscopic origin of CISP, SGE and SHE in the hopping regime is the spin-orbit interaction~\eqref{H_SO}. It results in precession of electron spins during the hops. The electron Hamiltonian in the basis of localized states has the form~\cite{KozubPRL}
\begin{equation}
 {\cal H}_e=\sum_{i,\sigma}\epsilon_ic_{i\sigma}^\dag c_{i\sigma}+\sum_{ij}\sum_{\sigma\sigma'}{J}_{ij}^{\sigma\sigma'}c_{i\sigma}^\dag c_{j\sigma'}.
\label{eq:ham}
\end{equation}
Here $c_{i\sigma}^\dag(c_{i\sigma})$ are the creation (annihilation) operators of an electron at the site $i$ with the spin projection ${\sigma = \pm 1/2}$ on the normal to the 2D plane, $z$ axis. The site energies consist of three contributions: ${\epsilon_i=E_b+U_i-e\bm E \cdot \bm R_i}$, where $E_b$ is the binding energy assumed to be equal for all sites, $U_i$ is the fluctuating electrostatic potential energy at the site,
and the last term describes the potential in the external electric field for the site with the 2D coordinate $\bm R_i$. The energies $U_i$ are broadly distributed, 
 and the variable-range hopping regime is realized~\cite{Efros89_eng}. The second term in Eq.~\eqref{eq:ham} describes spin-dependent hopping with the amplitudes~\cite{KKavokin-review,KozubPRL}
\begin{equation}
\hat{J}_{ij}=J_{ij}\e^{-\i\bm{d}_{ij} \cdot \bm{\hat{\sigma}}},
\quad
\bm{d}_{ij}=m \hat{\bm \beta}(\bm{R}_i-\bm{R}_j)/\hbar^2,
\end{equation}
where $J_{ij}$ are spin-independent hopping amplitudes between sites $i$ and $j$, and $m$ is the electron effective mass.

\textit{Kinetic equation.}---Electron transport in the studied spin-orbit coupled system is described by a kinetic equation for the spin density matrix. Decomposing the on-site density matrix as $\hat{\rho}_i=n_i/2+\hat{\bm{\sigma}}\cdot \bm S_i$, 
we derive a system of coupled equations for the site occupations $n_i$ and the spins $\bm S_i$~\cite{BryksinKinetiks,Supp}:
  \begin{subequations}
  \begin{equation}
    \dot{n}_i=\sum_j I_{ij}+\sum_{j}\left(\bm{\Lambda}_{ij} \cdot \bm{S}_j-\bm{\Lambda}_{ji} \cdot \bm{S}_i\right),
\label{eq:kinetic_n}
  \end{equation}
%
\begin{equation}
\dot{\bm S}_i +\sum_j \bm{S}_{j}\times \bm{\Omega}_{ij}+\frac{\bm S_i}{\tau_s} = \sum_{j} \bm I_{ij}^s +
\sum_{j} \left({\bm G}_{ij} n_j+{\bm G}_{ji} n_i\right).
\label{eq:kinetic_s}
  \end{equation}
\end{subequations}
Here 
$I_{ij}=n_j/\tau_{ij}-n_i/\tau_{ji}$ is the particle flow between sites $i$ and $j$
with $\tau_{ji}$ being the hopping time from the site $i$ to the site $j$. The second sum in Eq.~(\ref{eq:kinetic_n}) represents the source of an electric current induced by a nonequilibrium spin polarization, being the precursor of SGE.

The left hand side of Eq.~\eqref{eq:kinetic_s} has the form of Bloch equation with the effective frequency of spin precession during the hop
$
\bm{\Omega}_{ij}=2\bm{d}_{ij}/\tau_{ij},
$
%
and the on-site phenomenological spin relaxation time $\tau_s$, caused by the hyperfine interaction. This time is 
shorter than Dyakonov-Perel spin relaxation time in the hopping regime~\cite{Lyubinskiy}, and for the sake of simplicity we neglect the possible non-exponential spin relaxation dynamics. The spin current flowing from the site $j$ to the site $i$, ${\bm{I}}^s_{ij}$, is a sum of two contributions
\begin{equation}
{\bm{I}}^s_{ij}=\frac{\bm{S}_j}{\tau_{ij}}-\frac{\bm{S}_i}{\tau_{ji}}
	+ \bm{W}_{ij} n_j-\bm{W}_{ji} n_i.
\label{eq:Is}
\end{equation}
The first two terms describe the spin diffusion, while the latter terms arise due to a difference in spin-conserving tunneling rates for electrons with spin oriented along ($\uparrow$) and opposite ($\downarrow$) to the axis $\alpha$: $ W_{ij}^\alpha=(W_{\uparrow\uparrow}-W_{\downarrow\downarrow})/2$. This contribution clearly leads to a spatial separation of electrons with opposite spins $s_z=\pm1/2$ in the static electric field, which is a \textit{dc} SHE.

The last term in Eq.~(\ref{eq:kinetic_s}) describes the spin generation. It can be expressed via a difference of spin-flip probabilities during the hops
as  $G_{ij}^\alpha=\left(W_{\uparrow\downarrow}-W_{\downarrow\uparrow}\right)/2$.
We note that
the kinetic coefficient $\Lambda^\alpha_{ij}$ can also be presented as $2\left(W_{\uparrow\uparrow}+W_{\downarrow\uparrow}-W_{\downarrow\downarrow}-W_{\uparrow\downarrow}\right)$, thus allowing to find a fundamental relation between the kinetic coefficients
  \begin{equation}
     	\bm{\Lambda}_{ij} = 4\left(\bm{W}_{ij}-\bm{G}_{ij}\right).
        \label{eq:relation}
  \end{equation}

At the microscopic level, the spin dependence of the tunneling rates appears due to an interference of the direct hopping path with the hopping through an auxiliary site~\cite{Holstein,KozubPRL}. An arbitrary triad of localization sites is shown in the center of Fig.~\ref{fig:sketch}. 
The matrix element
of tunneling between the sites $1$ and $3$ up to the second order in hopping amplitude equals to
\begin{multline}
  \hat{J}_{31}+\frac{\hat{J}_{32}\hat{J}_{21}}{\Delta E_{12}}= \\ \hat{J}_{31}\left(1+\frac{J_{32}J_{21}}{J_{31}\Delta E_{12}}\e^{\i\bm{d}_{31}\cdot\hat{\bm{\sigma}}}\e^{-\i\bm{d}_{32}\cdot\hat{\bm{\sigma}}}\e^{-\i\bm{d}_{21}\cdot\hat{\bm{\sigma}}} \right),
\end{multline}
where $\Delta E_{12}$ is the energy difference between states $1$ and $2$, including the phonon energy.
Due to noncommutativity of Pauli matrices, the second term in the brackets is not reduced to a scalar: The electron spin orientation is changed after a travel over the closed path. This is due to the Berry curvature~\cite{Manchon2015,Sinova2015} arising from the inversion symmetry breaking in hopping Hamiltonian~\cite{Haldane}. As a result, the hopping matrix element is essentially spin dependent. Therefore the kinetic
 coefficients $\mathcal{K}_{ij}$ ($\mathcal{K}=\Lambda, G,W$) can be presented as a sum over the auxiliary sites
${\mathcal{K}_{ij}=\sum\limits_k\mathcal{K}_{ikj},}$
and the relation~\eqref{eq:relation} holds for $\mathcal{K}_{ikj}$ as well.
%
%
Microscopic calculation yields the kinetic coefficients~\cite{Supp}
  \begin{subequations}
  \begin{equation}
    \bm{G}_{ikj}=3Q_{ikj}\bm{A}_{ikj}\times\hat{\bm \beta}\bm{R}_{ij}\Tr\hat{\bm \beta}^2,
  \end{equation}
  \begin{equation}
    \bm{W}_{ikj}=Q_{ikj}\Tr\hat{\bm \beta}^2\left[\bm{A}_{ikj}\times\hat{\bm \beta}\left(\bm{R}_{jk}+\bm{R}_{ik}\right)
    -3\frac{\hbar^2}{m}\bm{A}_{ikj}\right],
  \end{equation}
  \end{subequations}
where $\bm{R}_{ij}=\bm{R}_i-\bm{R}_j$, $\bm{A}_{ikj}=\bm{R}_{ki}\times\bm{R}_{ij}/2$ is the oriented area of the triad, and $Q_{ikj}$ is the constant determined by the hopping times and hopping amplitudes between the sites~\cite{Supp}.
We note that the spin separation ($W_{ikj}^z$) appears in the second order in spin-orbit interaction, while the spin generation rate and spin galvanic current are cubic in the spin splitting.

\begin{figure}[t]
\centering\includegraphics[width=\linewidth]{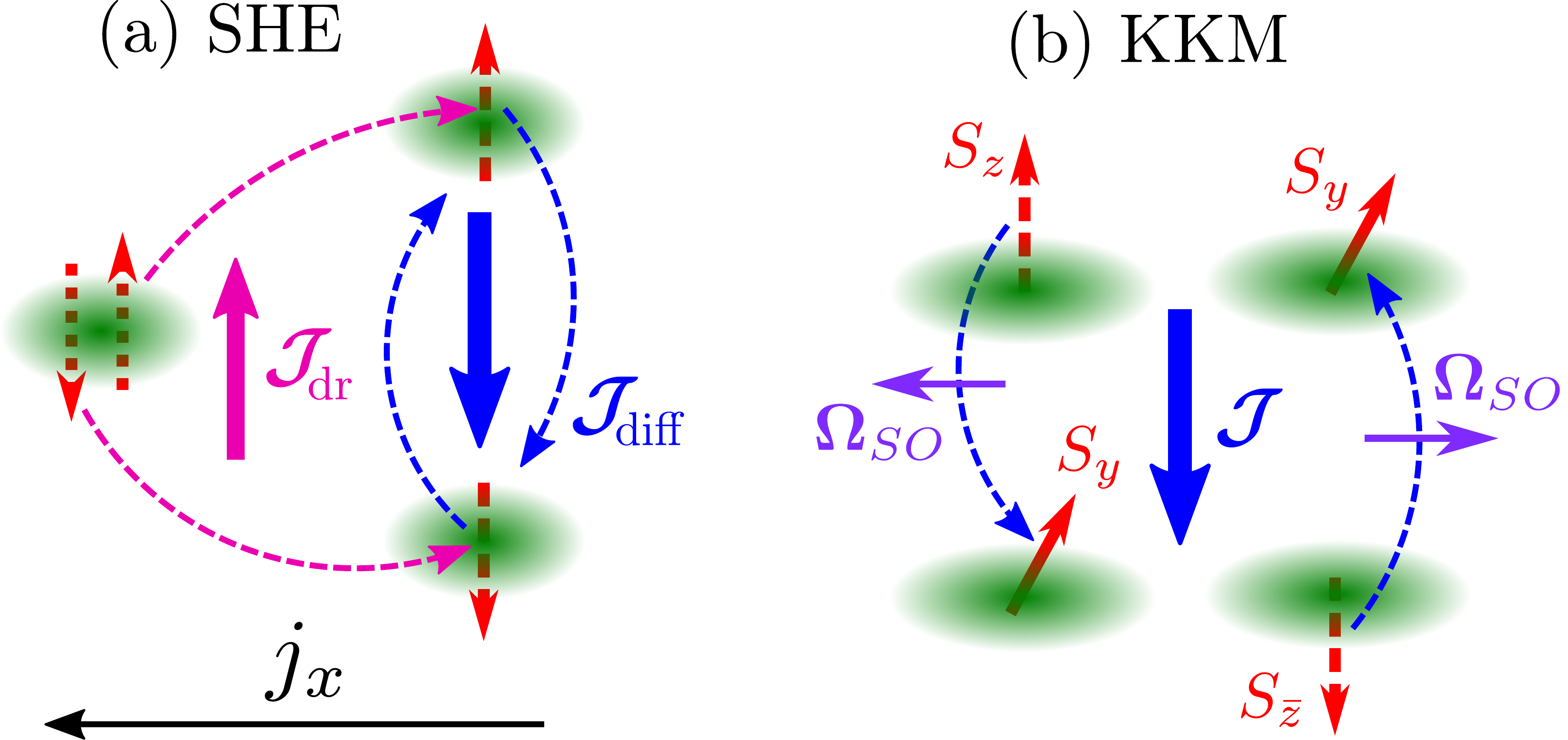}
\caption{Microscopic mechanism of CISP. (a) Electric current flow leads to the drift spin current, $\bm{\mathcal{J}}_{\rm dr}$, due to the SHE. In the strongly inhomogeneous system under study, it is partially compensated by the diffusion spin current, $\bm{\mathcal{J}}_{\rm diff}$. (b) Spin current, $\bm{\mathcal J}$, accompanied by the spin precession in spin-orbit field $\bm\Omega_{SO}$, results into electron spin polarization $S_y$ due to the KKM effect.}
\label{fig:CISP}
\end{figure}

\textit{Results.}---The CISP and SGE can be conveniently related to the spin current, $\Js$, flowing in the system.  
Indeed, the electric current leads to generation of the spin current due to the SHE, Fig.~\ref{fig:CISP}(a).
Then, the spin current is converted to spin polarization.
The effects of mutual spin and spin current conversion were introduced for free electrons in Ref.~\cite{KKM} by Kalevich, Korenev and Merkulov, and can be referred to as the KKM effects. 
For localized carriers it is illustrated in Fig.~\ref{fig:CISP}(b): In the presence of spin current, spin-up electrons ($S_z$) and spin-down electrons ($S_{\bar{z}}$) hop in opposite directions, and experience spin precession with frequency $\Omega_{SO}$ in opposite directions, which leads to spin polarization $S_y$.
Formally the KKM effect in the hopping regime can be derived from Eq.~\eqref{eq:kinetic_s} by taking the sum over all sites: 
  \begin{equation}
     {\bm s} = -\frac{2\tau_sm}{n\hbar^2}\bm e_z\times\hat{\bm\beta}{\bm{\mathcal{J}}},
\label{eq:IKKM}
  \end{equation}
where the spin current is defined as~\cite{KozubPRL,Raimondi2009}
\begin{equation}
  \Js=\frac{1}{2}\sum_{ij}{\bm R}_{ij}I_{ij}^{s,z},
\label{eq:Js}
\end{equation}
and $\bm e_z$ is a unit vector along the $z$ axis.

\begin{figure}[t]
\centering\includegraphics[width=\linewidth]{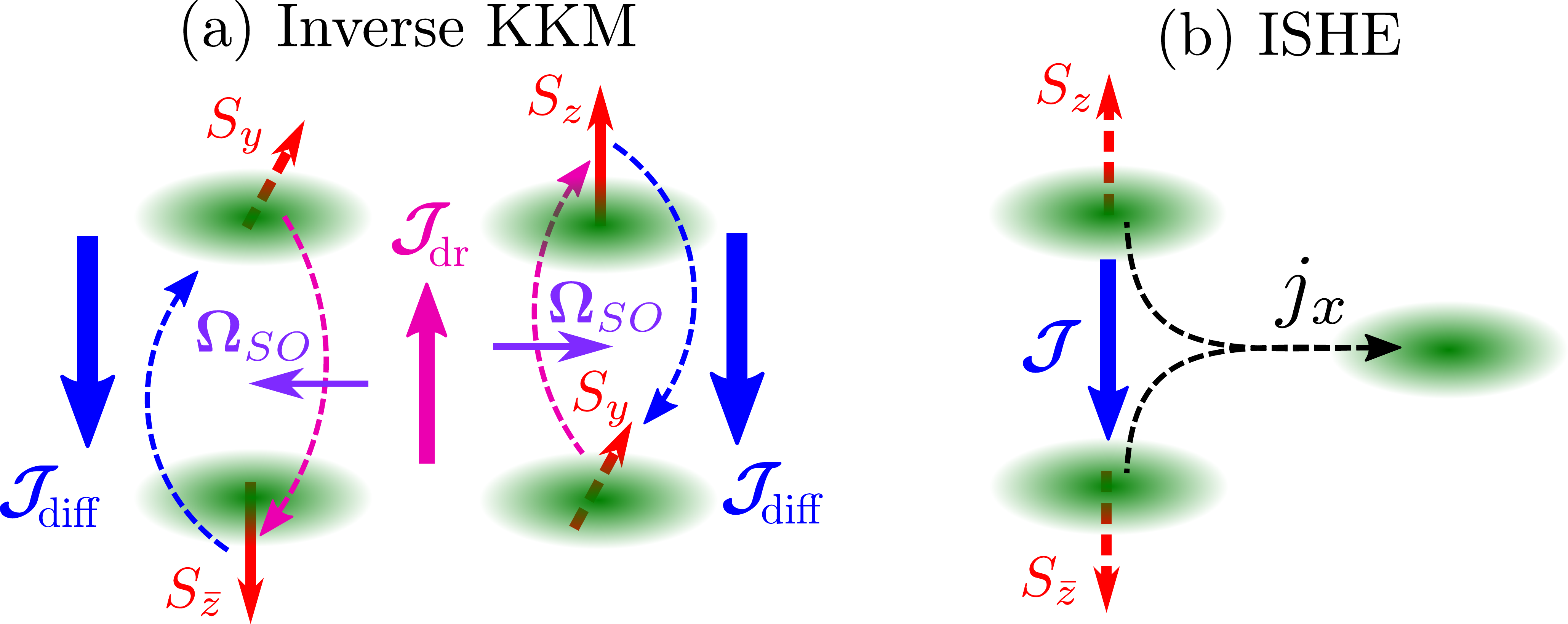}
\caption{Mechanism of SGE. (a) Spin polarization leads to the drift spin current, $\bm{\mathcal{J}}_{\rm dr}$,  due to the inverse KKM effect. It is partially compensated by the diffusion spin current, $\bm{\mathcal{J}}_{\rm diff}$. (b) Spin current $\bm{\mathcal J}$ results in the electric current $j_x$  due to ISHE.}
\label{fig:SGE}
\end{figure}

Spin-galvanic effect can be treated in a similar way, see Fig.~\ref{fig:SGE}. Spin polarization $S_y$ due to the spin-orbit interaction leads to the spin current, $\Js$ (inverse KKM effect). In turn, the spin current induces the electric current $j_x$ due to the inverse spin-Hall effect (ISHE). Therefore both CISP and SGE are intimately related to the spin current and can be decomposed into two steps, SHE+KKM and inverse KKM + inverse SHE, respectively. In fact CISP and SGE are reciprocal to each other due to time reversal symmetry~\cite{Gorini}.

\begin{figure}
  \includegraphics[width=\linewidth]{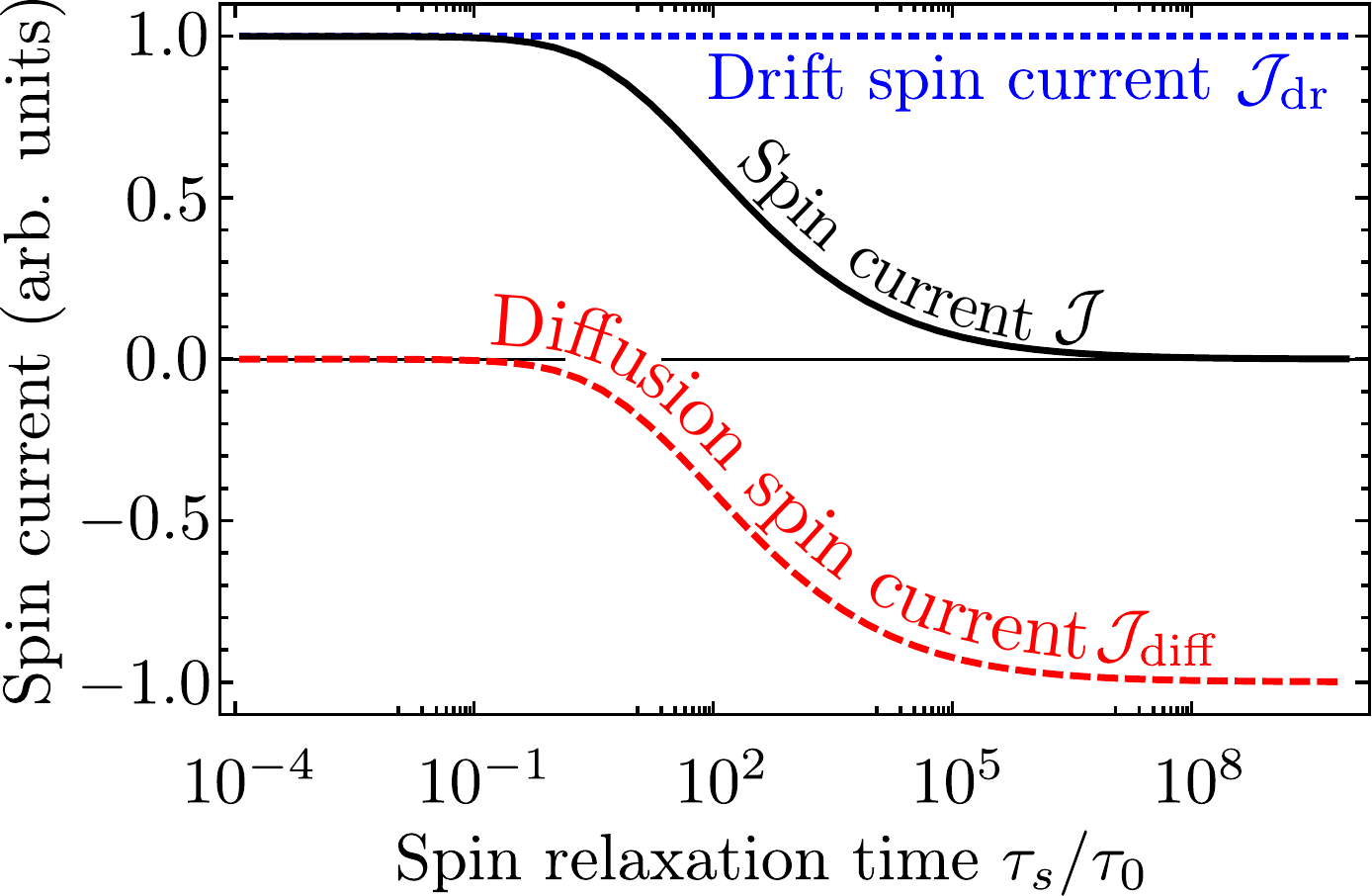}
  \caption{Drift, diffusion and total spin currents as functions of hyperfine induced spin relaxation time for $n_sa_b^2=0.01$. The black curve shows the function $f(n_s,\tau_s)$.}
  \label{fig:Js}
\end{figure}

The spin current defined by Eq.~\eqref{eq:Js} consists of two contributions: diffusion spin current, $\Jdiff$, and drift spin current, $\Jdr$, which correspond to the two terms of Eq.~\eqref{eq:Is}. Since the system under study is strongly inhomogeneous, the drift spin current leads to spin separation in the steady state. This, in turn, induces the diffusion spin current in the opposite direction as shown in Fig.~\ref{fig:CISP}(a). Neglecting the spin relaxation these two contributions completely cancel each other, so $\Js=\Jdiff+\Jdr$ is zero. The on-site hyperfine induced spin relaxation diminishes spin separation 
and upsets the balance, therefore
  \begin{equation}
    \Js=\frac{1}{\tau_s}\sum_i{\bm R}_iS_i^z.
  \end{equation}
This expression shows that in the limit of infinite nuclei-induced spin relaxation time the total spin current vanishes, but CISP has a finite value, see Eq.~\eqref{eq:IKKM}. The interplay between drift and diffusion spin currents is illustrated in Fig.~\ref{fig:Js}. Usually $\tau_s$ is much longer than the characteristic hopping time, $\tau_{ij}$, so the total spin current is less than both ${\mathcal{J}}_\text{dr}$ and ${\mathcal{J}}_\text{diff}$.

The hopping amplitude exponentially decreases with the increase of a distance between the sites, $J_{ij}\sim\exp\left(-R_{ij}/a_b\right)$, where the localization radius $a_b$ is assumed to be the same for all sites. This
gives an opportunity to make a quantitative analysis of the spin effects in the hopping regime where $n_s a_b^2 \ll 1$. To that end we extend the percolation theory~\cite{Efros89_eng}
to account for spin degrees of freedom. The electric current in the hopping regime flows only in the so-called percolation cluster, where the distances between the sites are 
the smallest,
and the potential energies are close to each other. The current induced spin polarization takes place only in the vicinity of this path.
%
The interference between the hopping paths also drops rapidly down at the distances larger than $a_b$. Since the electric current is the same in the whole cluster,
the main contribution to spin generation is given by the smallest triads of sites having the size $\sim a_b$~\cite{close}. Therefore the CISP conductivity
can be presented as~\cite{Supp}
\begin{equation}
\label{sigma_CISP}
  \hat{\bm \sigma}_\text{CISP}= \tau_s \Tr(\hat{\bm\beta}^2) \hat{\bm \beta}^T\mathcal{P}f(n_s,\tau_s).
\end{equation}
Here $\mathcal{P}=\left({m a_b}/{\hbar^2}\right)^3\hbar n_sa_b/(enJ_0\tau_0\rho)$, $\rho$ is the resistivity, $J_0$ and $\tau_0$ are the characteristic hopping integral and time for the distance $\sim a_b$, and $f(n_s,\tau_s)$ is a dimensionless function which tends to a finite value as $n_s$ goes to zero. 

The spin-galvanic current can be similarly obtained from the kinetic equation~\eqref{eq:kinetic_n}. It is generated also in small triads of sites and flows mainly in the percolation cluster. The calculation yields the following result for the SGE response~\cite{Supp}:
  \begin{equation}
      \hat{\bm \sigma}_\text{SGE}=4 \Tr(\hat{\bm\beta}^2) \hat{\bm \beta}^T \mathcal{P}k_BT n f(n_s,\tau_s).
  \end{equation}
Here the function $f$ coincides with that for CISP, Eq.~\eqref{sigma_CISP}, see Supplemental Material~\cite{Supp}. This coincidence comes from
the Onsager relation taking place for 
CISP and SGE susceptibilities due to reciprocity of these two 
effects~\cite{Levitov,Vignale}. We have analytically calculated the function $f(n_s,\tau_s)$ for the model of a regular triangle~\cite{Supp}.

The spin-Hall conductivity can be deduced from Eqs.~\eqref{eq:IKKM} and~\eqref{sigma_CISP}:
  \begin{equation}
    \hat{\bm\sigma}_\text{SHE}= -\hat{\bm\beta}^T \left(\bm{e}_z\times\hat{\bm\beta} \right)\frac{\hbar^2n\mathcal{P}}{m}f(n_s,\tau_s).
  \end{equation}
We stress that, in strongly inhomogeneous systems, the drift spin current is always accompanied by the diffusion spin current, and therefore the spin-Hall conductivity relates the applied electric field with the \textit{total} spin current, $\Js$. This conductivity 
 vanishes in the absence of hyperfine induced spin relaxation. However the spin separation is caused only by the drift spin current, which does not depend on spin relaxation, and therefore can be found as
  \begin{equation}
      \Jdr=-\hat{\bm\beta}^T\left(\bm{e}_z\times\hat{\bm\beta}\bm{E}\right)\frac{\hbar^2n\mathcal{P}}{m}f(n_s,0).
      \label{eq:Jdr}
  \end{equation}

The possibility of intrinsic spin current and spin separation for free electrons was intensively debated, for review see e.g. Refs.~\cite{Schliemann2006,Sinova2015}. It was found that the intrinsic spin-Hall effect is possible at the edges of the sample~\cite{Adagideli2005} or in mesoscopic systems~\cite{Nikolic2005}. In the hopping regime the electric current flows in a narrow quasi-one-dimensional cluster. Therefore the intrinsic spin current is expected to be nonzero in the strongly inhomogeneous system under study.

In order to make an estimation of CISP by Eq.~(\ref{sigma_CISP}) we present the hopping resistivity as $\rho=\rho_0\exp{(2l_c/a_b)}$, where $l_c$ is the maximum distance between neighboring sites in the percolation cluster. 
We adopt the model where the hopping amplitudes are ${J_{ij}=J_0\exp(-R_{ij}/a_b)}$ and $\tau_{ij}=\tau_0\exp(2R_{ij}/a_b)$. Under these assumptions $l_c\approx1.2/\sqrt{n_s}$. The numerical simulation of spin dynamics in the hopping regime was performed on the square sample with $5\times10^5$ sites~\cite{Supp}.

All the three effects under study are described by a single dimensionless function $f$ and obey the common dependence on the site concentration and on the spin relaxation time. The spin current as a function $\tau_0/\tau_s$ is shown in Fig.~\ref{fig:Js} for the small concentration $n_sa_b^2=0.01$. The drift and diffusion contributions to the spin current are shown separately. As expected, in the limit of slow spin relaxation the drift and diffusion currents completely compensate each other, so the total spin current is zero. We stress that the spin separation is still present in this limit and represents the intrinsic SHE. As the spin relaxation rate increases the diffusion spin current diminishes, and in the limit ${\tau_s=0}$ only the drift spin current survives in agreement with Eq.~\eqref{eq:Jdr}. For small concentrations $n_sa_b^2<0.02$ we find a finite value $f(n_s,0)\approx 1.0$, so the drift spin current is independent of $n_s$ in this limit.

For typical parameters  $n_s=20\,n=2\times 10^{11}$~cm$^{-2}$, $a_b=10$~nm, $J_0=10$~meV, $\beta_{xy}=\beta_{yx}=10$~meV\AA, $m=0.1m_0$, $E=1$~kV/cm, $\rho_0=50$~kOhm, $\tau_0=10$~ps, and ${\tau_s=100}$~ns, we obtain a small value $s\sim3\times 10^{-5}$. However, increase of $n_s$ results in drastic enhancement of the electrical spin polarization. At the crossover from hopping to the diffusion conductivity
we get a relatively large value of CISP $s \sim 1$~\% easily detectable in experiments.

\textit{Conclusion.}---We have proposed a unified description of CISP, SGE and SHE in the hopping regime. Based on numerical simulations and percolation theory we made the estimations of the corresponding susceptibilities. Due to the suppression of spin relaxation in the hopping conductivity regime, the spin effects are underlined, in particular the degree of current induced spin polarization for real structures can be large.

We thank M.\,M. Glazov and E.\,L. Ivchenko for fruitful discussions. Partial  support from 
RFBR (project 16-02-00375), the Dynasty Foundation and RF President Grant No. SP-643.2015.5
is gratefully acknowledged.


%






\onecolumngrid
\vspace{\columnsep}
\begin{center}
\newpage{\large\bf {Supplemental Material to\\ ``Electrical spin orientation, spin-galvanic and spin-Hall effects\\ in disordered two-dimensional systems''}
}
\end{center}
\vspace{\columnsep}
\twocolumngrid


\renewcommand{\cite}[1]{{[}\onlinecite{#1}{]}}

\renewcommand{\thepage}{S\arabic{page}}
\renewcommand{\theequation}{S\arabic{equation}}
\renewcommand{\thefigure}{S\arabic{figure}}
\renewcommand{\bibnumfmt}[1]{[S#1]}
\renewcommand{\citenumfont}[1]{S#1}
\setcounter{page}{1}
\setcounter{section}{0}
\setcounter{equation}{0}
\setcounter{figure}{0}

\section{S1. Derivation of kinetic equation for hopping regime}


The total Hamiltonian of the system reads
\begin{equation}
  {\cal H}={\cal H}_e+{\cal H}_{ph}+{\cal H}_{e-ph}.
\end{equation}
The electron Hamiltonian, ${\cal H}_e$ is given by Eq.~(3) of the main text. The phonon Hamiltonian is
\begin{equation}
   {\cal H}_{ph}=\sum_{\bm q}\hbar\Omega_{q}b_{\bm q}^\dag b_{\bm q},
\end{equation}
where $\hbar\Omega_{\bm q}$ is the energy of the phonon with the wavevector $\bm q$, and $b_{\bm q}(b_{\bm q}^\dag)$ is its annihilation (creation) operator~\cite{q}. The Hamiltonian of the electron-phonon interaction reads
\begin{equation}
   {\cal H}_{e-ph}=\sum_{i,\sigma,\bm{q}}v_q(\e^{\i\bm{qR_i}}b_{\bm q}+\e^{-\i\bm{qR_i}}b_{\bm q}^\dag)c_{i\sigma}^\dag c_{i\sigma}
\end{equation}
with $v_q$ being the electron-phonon interaction constants.

After the canonical transformation~\cite{Polarons,BryksinReview}, the Hamiltonian can be presented as
\begin{equation}
  {\cal H}=\sum_{i,\sigma}\epsilon_ic_{i\sigma}^\dag c_{i\sigma}+\sum_{\bm q}\hbar\Omega_qb_{\bm q}^\dag b_{\bm q}+\sum_{i,j,\sigma,\sigma'}{V}_{ij}^{\sigma\sigma'}c_{i\sigma}^\dag c_{j\sigma'},
\end{equation}
where $V_{ij}^{\sigma\sigma'}={J}_{ij}^{\sigma\sigma'}Q_{ij}$ with
\begin{equation}
  Q_{ij}=\exp\left\lbrace-\sum_{\bm q}\gamma_q\left[\left(\e^{\i\bm{qR}_i}-\e^{\i\bm{qR}_j}\right)b_{\bm q}+\rm{h.c}\right]\right\rbrace,
\label{eq:Vij}
\end{equation}
and $\gamma_q=v_q/(\hbar\Omega_q)$.

We decompose the density matrix of the electron system into a direct product of on-site density matrices, $\rho_i$. In the second order of the perturbation theory in hopping amplitudes, the time derivative of $\rho_i$ reads:
\begin{align}
\label{eq:rho2}
 & \dot{\rho}_i=\frac{\pi}{\hbar} \\
	\times & \left\langle\delta(E_n-E_m)\left(2V_{nm}\rho_jV_{mn}-\rho_iV_{nm}V_{mn}-V_{nm}V_{mn}\rho_i\right)\right\rangle. \nonumber
\end{align}
Here $n$ and $m$ denote the states of the electron-phonon system, where the given electron is localized at sites $i$ and $j$, respectively. The angular brackets denote averaging over the phonon bath state.

The hopping time $\tau_{ij}$ can be calculated from Eq.~\eqref{eq:rho2} as
\begin{equation}
  \frac{1}{\tau_{ji}}=\frac{2\pi}{\hbar}\Tr_{ph}\left[\rho_{ph}\hat{V}_{nm}\hat{V}_{mn}\delta(E_n-E_m)\right],
\label{eq:tauV}
\end{equation}
where $\rho_{ph}$ is the phonon density matrix, and the trace is taken over the phonon degrees of freedom.

\begin{figure}
  \centering\includegraphics[width=0.9\linewidth]{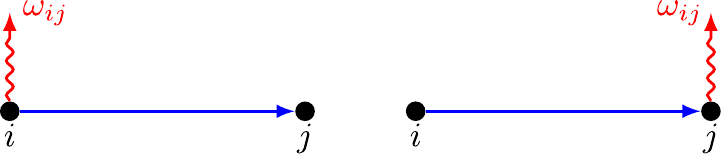}
  \caption{Illustrations of the second order contributions to the hopping probability. The hops are shown by blue arrows, while phonon emission is illustrated by red wavy arrows. It is assumed that $\epsilon_i>\epsilon_j$ and $\hbar\omega_{ij}=\epsilon_i-\epsilon_j$; in case of $\omega_{ij}<0$ the phonon lines should be reversed.}
  \label{fig:2order}
\end{figure}

 Since
\begin{equation}
  \hat{V}_{nm}\hat{V}_{mn}=J_{ij}^2\hat{I} Q_{ij}Q_{ji},
\end{equation}
the time $\tau_{ji}$ is the same for all spin orientations. In the lowest (second) order in the electron-phonon interaction  the hopping time is given by
\begin{equation}
  \frac{1}{\tau_{ji}}=
\frac{2\pi}{\hbar}J_{ij}^22\gamma_{q_{ij}}^2D(\left|\epsilon_{ij}\right|)\left(N_{|\epsilon_{ij}|}+\Theta(\epsilon_{ij})\right),
\label{eq:tauij}
\end{equation}
where $\epsilon_{ij}=\epsilon_i-\epsilon_j$, $q_{ij}$ is the phonon wave vector corresponding to this energy, $\Theta(\epsilon)$ is the Heaviside function, $D(\epsilon)$ stands for the density of phonon states, and $N_\epsilon=1/\left[\exp(\epsilon/k_B T)-1\right]$ is the occupation of the phonon state. The multiplier $2$ reflects the fact that the phonon can be emitted either at site $i$ or $j$, as shown in Fig.~\ref{fig:2order}. Note that $\tau_{ji}\neq\tau_{ij}$ despite the apparent symmetry of Eq.~\eqref{eq:tauV}. The two and more electron hops are disregarded in this work for the sake of simplicity.

In order to derive the spin generation rate one has to consider the third order in hopping. The corresponding generalization of Eq.~\eqref{eq:rho2} to this case reads~\cite{density_LL,density_KohnLutt,density_Sturman,density_Glazov}
\begin{multline}
 \dot{\rho}_i=-\frac{4\pi}{\hbar}\biggl\langle \delta(E_n-E_m) \biggl\{
\pi\Im\left(V_{nm}\rho_jV_{ml}V_{ln}\right)\delta(E_n-E_l)
\\ 
+{\Re\left[\rho_i\Re\left(V_{nm}V_{ml}V_{ln}\right) - V_{nm}\rho_jV_{ml}V_{ln}\right] \over E_n-E_l}
\biggr\} \biggr\rangle,
\label{eq:rho3}
\end{multline}
where the given electron is localized at the site $i,j$ or $k$ in the state $n,m$ and $l$, respectively, and we have introduced the notations
\begin{equation}
  \Re A=\frac{A+A^\dag}{2}, \quad
  \Im A=\frac{A-A^\dag}{2\i}.
\end{equation}
Clearly 
second line in Eq.~\eqref{eq:rho3} describes spin-independent hopping between the sites $i$ and $j$, while the 
first line describes a dependence of hopping probability on spin.
Note that this term does not contain $\rho_i$, i.e. it is pure ingoing term, so the outgoing terms are spin-independent.

The spin generation, spin separation and spin galvanic effects are described by the 
first line of Eq.~\eqref{eq:rho3} which can be rewritten as
\begin{multline}
 \i\frac{4\pi^2}{\hbar}\left\langle V_{nm}\rho_jV_{ml}V_{ln}\delta(E_n-E_m)\delta(E_n-E_l)\right>
\\
  \equiv\i\frac{4\pi^2}{\hbar} \hat{J}_{ij}\rho_j\hat{J}_{jk}\hat{J}_{ki}{\cal Q}_{ikj},
\end{multline}
where 
\begin{equation}
  {\cal Q}_{ikj}=\Tr_{ph}\left[Q_{ij}\rho_{ph}Q_{jk}Q_{ki}\delta(E_n-E_m)\delta(E_m-E_l)\right].
\label{eq:phonons3}
\end{equation}

The absence of spin dependence in the outgoing terms of
Eq.~\eqref{eq:rho3} means that (c.f. Eqs.~(5b) and~(6) of the main text)
\begin{equation}
  \bm W_{jki}+\bm W_{kji}=\bm G_{jki}+\bm G_{kji}.
\end{equation}
Clearly the coefficient for the spin galvanic effect is
\begin{equation}
  \bm \Lambda_{ikj}=\i\frac{4\pi^2}{\hbar}\Tr\left(\hat{J}_{ij}\bm\sigma_\alpha \hat{J}_{jk}\hat{J}_{ki}\right){\cal Q}_{ikj}.
\label{eq:lambda}
\end{equation}
The spin generation rate at the given site, $\bm \Gamma_{ikj}$, consists of two contributions, namely spin generation and spin separation: $\bm\Gamma_{ikj}\equiv \bm W_{ikj}+ \bm G_{ikj}$. It can be found as
\begin{equation}
  \bm\Gamma_{ikj}=\i\frac{\pi^2}{\hbar}\Tr\left(\bm\sigma \hat{J}_{ij}\hat{J}_{jk}\hat{J}_{ki}\right){\cal Q}_{ikj}.
\label{eq:gamma}
\end{equation}

\section{S2. Calculation of kinetic coefficients}

Expressions Eq.~\eqref{eq:lambda} and~\eqref{eq:gamma} combined with Eq.~(7) of the main text allow one to find
\begin{equation}
  \bm G_{ikj}=\i\frac{\pi^2}{2\hbar}\left[\Tr\left(\bm\sigma \hat{J}_{ij}\hat{J}_{jk}\hat{J}_{ki}\right)-\Tr\left(\hat{J}_{ij}\bm\sigma\hat{J}_{jk}\hat{J}_{ki}\right)\right]{\cal Q}_{ikj}.
\label{eq:generation}
\end{equation}
\begin{equation}
  \bm W_{ikj}=\i\frac{\pi^2}{2\hbar}\left[\Tr\left(\bm\sigma\hat{J}_{ij}\hat{J}_{jk}\hat{J}_{ki}\right)+\Tr\left(\hat{J}_{ij}\bm\sigma\hat{J}_{jk}\hat{J}_{ki}\right)\right]{\cal Q}_{ikj}.
\end{equation}
Up to the third order in the spin-orbit coupling these expressions are reduced to Eqs.~(9) of the main text with
\begin{equation}
  {Q}_{ikj}=\frac{2\pi^2m^3}{3\hbar^7}J_{ij}J_{jk}J_{ki}{\cal Q}_{ikj}.
\end{equation}


\begin{figure}
  \centering\includegraphics[width=0.9\linewidth]{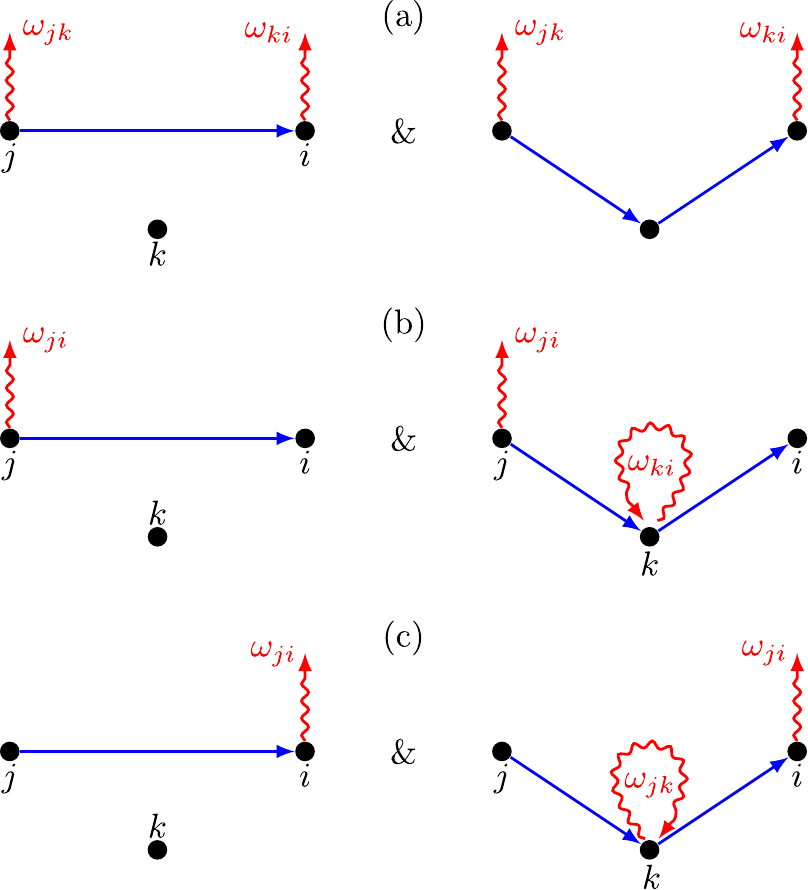}
  \caption{Illustrations of the lowest order interfering processes contributing to spin dependent hopping, $V_{mn}~\&~V_{ml}V_{ln}$. The notations are the same as in Fig.~\ref{fig:2order}.
}
  \label{fig:3order}
\end{figure}

The lowest order in the electron-phonon interaction contributions to Eq.~\eqref{eq:phonons3} are illustrated in Fig.~\ref{fig:3order}. One can see that the interference is possible between second-and-second [panel (a)] or first-and-third [panels (b) and (c)] orders in the electron-phonon interaction. Comparing Figs.~\ref{fig:2order} and~\ref{fig:3order}, one can easily deduce from  Eq.~\eqref{eq:phonons3}:
\begin{multline}
  {\cal Q}_{ikj}
  ={1\over 4}\bigl[
    \left\langle Q_{jk}Q_{kj}\delta(E_m-E_l)\right\rangle \left\langle Q_{ki}Q_{ik}\delta(E_n-E_l)\right\rangle\\
    +\left\langle Q_{ji}Q_{ij}\delta(E_m-E_l)\right\rangle \left\langle Q_{ik}Q_{ki}\delta(E_n-E_l)\right\rangle
    \\
    +\left\langle Q_{ji}Q_{ij}\delta(E_m-E_l)\right\rangle \left\langle Q_{jk}Q_{kj}\delta(E_n-E_l)\right\rangle
    \bigr].
\label{eq:QQQ}
\end{multline}
Finally 
we find
\begin{multline}
  Q_{ikj}=\frac{m^3}{24\hbar^5} \\
	\times \left(\frac{J_{ij}}{J_{kj}J_{ki}\tau_{ik}\tau_{kj}}+\frac{J_{jk}}{J_{ij}J_{ki}\tau_{ki}\tau_{ij}}+\frac{J_{ik}}{J_{ij}J_{kj}\tau_{ij}\tau_{kj}} \right).
\label{eq:Q}
\end{multline}

\section{S3. Analysis of kinetic equation}

\subsection{S3A. Thermal equilibrium}

In thermal equilibrium the populations of the sites and the phonon numbers obey Fermi-Dirac and Bose-Einstein distributions, respectively. Hence
\begin{equation}
  \frac{n_i}{n_j}=\e^{\epsilon_{ji}/k_BT},\qquad
  \frac{\tau_{ij}}{\tau_{ji}} = \frac{N_{|\epsilon_{ij}|}+\Theta(\epsilon_{ij})}{N_{|\epsilon_{ij}|}+\Theta(\epsilon_{ji})} = \e^{\epsilon_{ij}/k_BT}.
\label{eq:thermal}
\end{equation}
Using these relations
one can show that the particle flux is absent in thermal equilibrium:
\begin{equation}
\label{I_eq}
  I_{ij}= {n_j\over \tau_{ij}}-{n_i\over \tau_{ji}} = 0.
\end{equation}

Then, let us consider the spin generation rate in the given triad:
  \begin{equation}
    \bm{\Upsilon}_{ikj}\equiv \bm\Gamma_{ikj}n_j+\bm\Gamma_{ijk}n_k.
    \label{eq:upsilon1}
  \end{equation}
  One can show that in thermal equilibrium this expression also vanishes as expected. Interestingly the spin generation and spin separation contributions to $\bm{\Upsilon}_{ikj}$ do not vanish separately, which indicates the presence of persistent spin currents~\cite{S_KozubPRL}.



\subsection{S3B. Current induced spin generation}

Let us assume that the electric field was applied from the beginning, so the energies of the sites has been changed and the electrons have already redistributed to match thermal equilibrium. In the initial state the circuit was open and then we close it. In these conditions the populations of all the sites get the variation $\delta n_i$. 
The particle flux through an edge can be rewritten as
\begin{equation}
\label{I}
  I_{ij}=\left(\frac{\delta n_j}{n_j}-\frac{\delta n_i}{n_i}\right) I_{ij}^{(0)},
\end{equation}
where
\begin{equation}
\label{I0}
  I_{ij}^{(0)}=\frac{n_j}{\tau_{ij}}
\end{equation}
is the current of particles in one direction in thermal equilibrium, cf. Eq.~\eqref{I_eq}. Hereafter the populations, hopping times and kinetic coefficients are calculated in the equilibrium state.

Using Eqs.~\eqref{eq:upsilon1}---\eqref{I0}, the spin generation rate at the given site can be presented as
\begin{equation}
  \bm{\Upsilon}_{ikj}= \bm\Gamma_{ikj}\delta n_j+\bm\Gamma_{ijk}\delta n_k = I_{kj}\tau_{kj}\bm\Gamma_{ikj}.
\label{eq:upsilon}
\end{equation}
This expression directly links the spin generation with the electric current, but only for $x$ and $y$ spin components the average spin generation rate is nonzero.

It is interesting to analyze CISP in the limit of high site concentration, $n_sa_b^2\sim 1$, at the edge of our model applicability limit. In this case the Dyakonov-Perel spin relaxation time can be estimated as
  \begin{equation}
    \frac{1}{\tau_{DP}}=\frac{\varphi^2}{\tau_0},
  \end{equation}
where $\varphi\sim m\beta a_b/\hbar^2$ is the characteristic spin rotation angle during a hop. 
We obtain the simple estimation of the spin polarization in this limit
\begin{equation}
  s=\frac{\tau_s}{\tau_{DP}}\frac{\beta k_{\rm dr}}{J_0},
\end{equation}
where we have introduced a
``drift wave vector'', $k_{\rm dr}$, according to
\begin{equation}
  j=\frac{ne\hbar}{m}k_{\rm dr}.
\end{equation}
For the realistic parameters given in the main text the two spin relaxation times in this limit are, as expected, of the same order.


\section{S4. Relation between CISP and SGE susceptibilities}

In order to apply the Onsager relation~\cite{LL} for CISP and SGE we assume that the magnetic and electric fields oscillating at a low frequency $\omega$ are applied to the system. The spin and electromagnetic Hamiltonians, respectively, have the form
\begin{equation}
	H_S = n\bm s \cdot g\mu_B \bm B, \qquad H_{em} = - {1\over c} \bm j \cdot \bm A,
\end{equation}
where $\mu_B$ is the Bohr magneton, $\bm B$ is the magnetic field and $\bm A$ is the electromagnetic vector potential. 
The Onsager relation stands that the susceptibilities relating $n\bm s$ with $-\bm A/c$
and $\bm j$ with $g\mu_B \bm B$ should be equal~\cite{S_Levitov,S_Vignale}:
\begin{subequations}
  \begin{equation}
    n\bm s =  - {1\over c} \bm{\hat{\sigma}}_a  \bm A,
    \label{eq:OnsagerCISP}
  \end{equation}
  \begin{equation}
    \bm j = \bm{\hat{\sigma}}_a g\mu_B \bm B.
    \label{eq:OnsagerSGE}
  \end{equation}
\end{subequations}
From the first equation one immediately concludes that
\begin{equation}
  \label{eq:sigma_CISP}
  \bm\sigma_\text{CISP}=\bm{\hat{\sigma}}_a/(-\text{i}\omega n).
\end{equation}

The steady state spin polarization in magnetic field is
\begin{equation}
  \bm s_0=-\frac{g\mu_B {\bm B}}{4k_BT}.
\end{equation}
For the alternating magnetic field the spin polarization obeys equation
\begin{equation}
  \dot{\bm s}=\frac{\bm s_0-\bm s}{\tau_s}
\end{equation}
with the solution
\begin{equation}
  \bm s=\frac{\bm s_0}{1-\i\omega\tau_s} \approx -\frac{g\mu_B {\bm B}}{4k_BT}(1+\i\omega\tau_s)
	\equiv \bm s_0 + \delta \bm s.
\end{equation}
Using this expression one can rewrite Eq.~\eqref{eq:OnsagerSGE} as
\begin{equation}
  {\bm j}= \frac{4 k_B T}{\tau_s} {\bm{\hat{\sigma}}_a \over -\text{i}\omega} \delta\bm s.
\end{equation}
Therefore the SGE is described by
\begin{equation}
  \label{eq:sigma_SGE}
  \bm\sigma_\text{SGE}=\frac{4 k_B T}{\tau_s} {\bm{\hat{\sigma}}_a \over -\text{i}\omega}.
\end{equation}

Finally from comparison of Eqs~\eqref{eq:sigma_CISP} and~\eqref{eq:sigma_SGE} one finds
\begin{equation}
  \bm\sigma_\text{SGE}=\bm\sigma_\text{CISP} {4 k_B T \, n \over\tau_s}.
\end{equation}

\section{S5. Spatial disorder model}

\subsection{S5A. General analysis}

Microscopically one can distinguish two mechanisms of spin generation. The first one is a result of direct spin generation and is associated to drift spin current. The contribution to spin polarization from this mechanism is expressed as
\begin{equation}
    \bm s_{\rm dr}=\tau_s\frac{n_s}{n}\left\langle \sum_{kj}'\bm\Upsilon_{ikj}\right\rangle,
\end{equation}
where the stroke means that each pair $(kj)$ should be taken only once, without permutation.

An alternative way of spin generation is through the diffusion spin current~\cite{S_kkm94}. 
In this case spin separation $\pm s_z$ is converted to the spin polarization $s_y$. This mechanism can be in principle parametrically separated from the previous one e.g. in case of anisotropic spin relaxation. The contribution to spin polarization from this mechanism reads
\begin{equation}
    \bm s_{\rm diff}=\tau_s\frac{n_s}{n}\left\langle \sum_{j} \bm\Omega_{ij}\times \bm S_j\right\rangle.
\end{equation}

In the simplest model we assume that $J_{ij}=J_0\e^{-R_{ij}/a_b}$ and $\tau_{ij}=\tau_0\e^{2R_{ij}/a_b}\sim \rho_{ij}$, where $\rho_{ij}$ is the resistance of the edge between sites $i$ and $j$. In this case one can calculate the spin generation rates on the basis of Eq.~\eqref{eq:Q}. 
The contribution to the function $f(n_s,\tau_s)$ related with the drift spin current can be presented as
\begin{multline}
  f_{\rm dr}(n_s) =\frac{L}{8Na_b} \\
	\times {\sum\limits_{ijk}'I_{kj}
	(\bm R_{ij} \times \bm R_{ik})_z
	(R_{ij}^y+R_{ik}^y)\e^{\left(R_{kj}-R_{ik}-R_{ij}\right) /a_b} \over  {a_b^3}\sum\limits_{\rm{border}}I_{ij} },
\label{eq:fp}
\end{multline}
where the denominator represents the total particle flow through the sample, thus the sum should be taken only over one border. $N=n_sL^2$ is the number of localization sites in the sample with $L$ being its length. To be specific we have assumed that the current flows along $x$ axis.

In order to calculate the contribution from the second mechanism,  we first find normalized spins $\tilde S_i^z$.
To that end we solve the  equation
\begin{equation}
  \dot{\tilde S}_i^z = \sum_{jk} \gamma_{ikj} + \sum_j\left(\frac{\tilde S_j^z}{\tau_{ij}} - \frac{\tilde S_i^z}{\tau_{ji}}\right) - \frac{\tilde S_i^z}{\tau_s}
\end{equation}
with the normalized source
\begin{equation}
  \gamma_{ikj}= -I_{kj} \frac{3}{16a_b^2\tau_0} (R_{ij}^xR_{ik}^y-R_{ik}^xR_{ij}^y)\e^{\left(R_{kj}-R_{ik}-R_{ij}\right)/a_b}.
\end{equation}
Then we find the contribution to $f$ from precessional mechanism as
\begin{equation}
  f_{\rm diff}(n_s,\tau_s) = -\frac{2L}{Na_b} \left[\sum_{ij}\frac{R_{ij}^y}{a_b} \tilde S_j^z\e^{-2R_{ij}/a_b}\right] / \sum_{\rm{border}}I_{ij}.
\label{eq:fi}
\end{equation}

The fact that the function $f(n_s,\tau_s)$ calculated after Eqs.~\eqref{eq:fp} and~\eqref{eq:fi} is of the order of unity supports the reasoning that the main contribution to spin generation is given by small triads of sites.

\subsection{S5B. Elementary equilateral triangle}

\begin{figure}[h]
  \centering\includegraphics[width=0.9\linewidth]{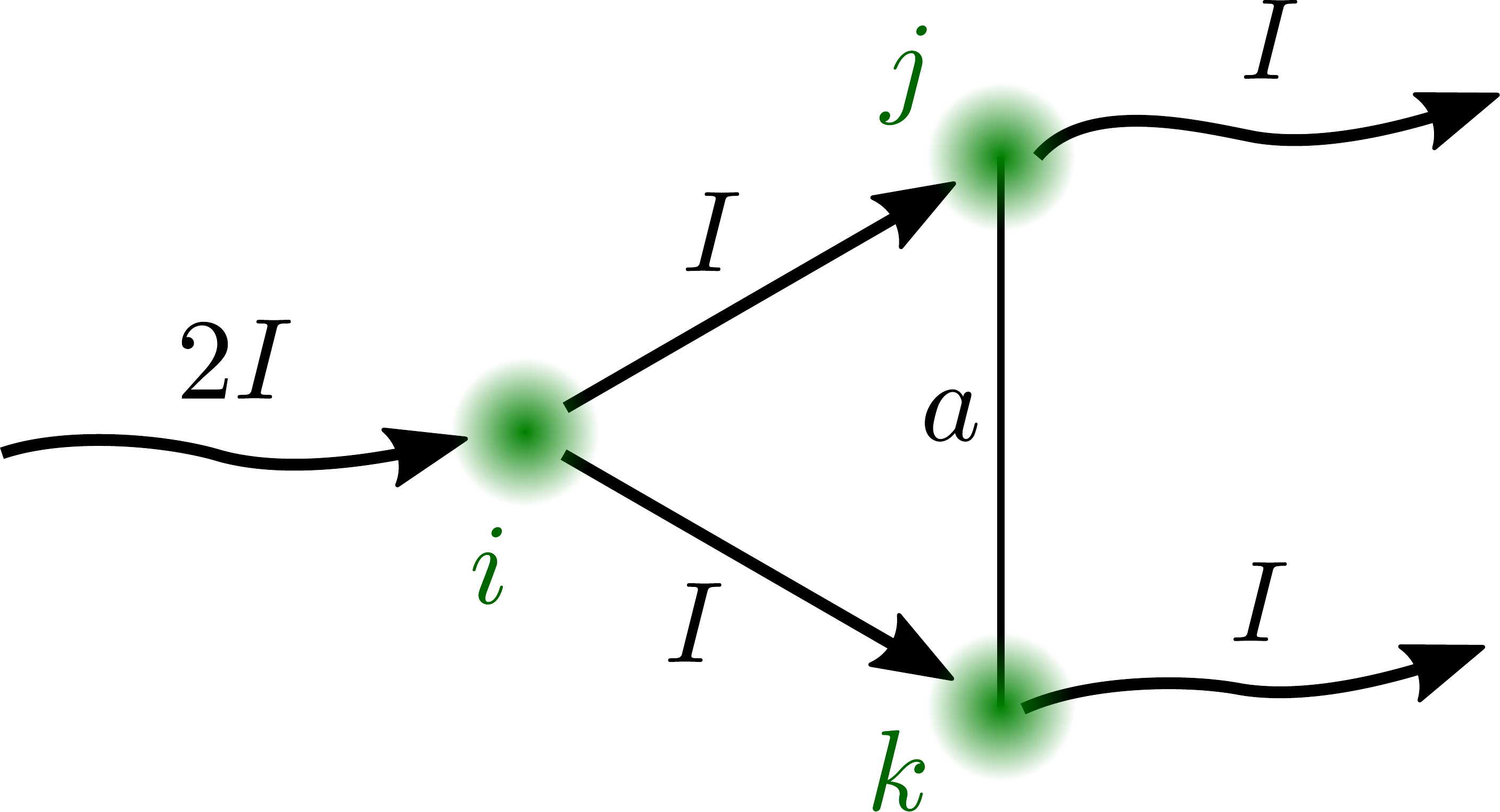}
  \caption{The simplest configuration of sites, which allows to compare two mechanisms of spin generation.}
  \label{fig:triangle}
\end{figure}

The simplest configuration of sites is an equilateral triangle, as shown in Fig.~\ref{fig:triangle}. Let the side of triangle be $a$. Explicit calculation after Eqs.~\eqref{eq:fp} and~\eqref{eq:fi} yields
\begin{equation}
  f_{\rm dr}=\frac{\sqrt{3}a^4}{32a_b^4}\e^{-a/a_b}, 
  \quad
  f_{\rm diff}=-f_{\rm dr}\frac{\tau_s}{\tau_s+(1/3)\tau_0\e^{2a/a_b}}.
\label{eq:Sp}
\end{equation}
As expected, in the limit $\tau_s\to\infty$ the two contributions cancel each other. We note that the same expressions can be also obtained from the calculation of SGE effect, in agreement with Onsager relation.




\subsection{S5C. Numerical simulation}

Numerically the function $f(n_s,\tau_s)$ was calculated after Eqs.~\eqref{eq:fp} and~\eqref{eq:fi}. In order to find the currents $I_{ij}$ we have solved the system of Kirchhoff equations on the resistivity network, $\rho_{ij}=\rho_0\tau_{ij}/\tau_0$.

We have limited ourselves to the electrical connections and hops of the length smaller than $1.5\,l_c$ only. This value is acceptable for the concentrations $n_sa_b^2<0.1$. The result of each calculation was tested on robustness by at least two other realizations, and for $N=5\times 10^5$ sites the deviations are smaller than 1\%.


\begin{thebibliography}{25}
\bibitem{QW110}
The presented theory can also be applied to asymmetric $(110)$ heterostructures, whos Hamiltonian is restricted to Eq.~\eqref{H_SO} after the coordinate frame rotation in the spin space.
\bibitem{close} The perturbation theory in hopping amplitude can not be applied for such small triads, and they should be analyzed separately. Nevertheless the presented expressions for the susceptibilities give the correct estimations of the effects.
\end{thebibliography}
\begin{thebibliography}{nofootinbib}
\bibitem{q} The index of phonon polarization is included in $\bm q$.
\end{thebibliography}
\end{document}